\documentclass[twocolumn,aps,showpacs,floatfix]{revtex4}

\usepackage{graphicx}
\usepackage{bm}
\begin{document}

\title{Collisional properties of ultracold K-Rb mixtures}

\author{G. Ferrari, M. Inguscio, W.
  Jastrzebski\thanks{also at Institute of Physics, Polish Academy of
    Science, 02668 Warszawa, Poland}, G. Modugno, G. Roati\thanks{also
    at Dipartimento di Fisica, Universit\'a di Trento, 38050 Povo,
    Italy}} \address{LENS and Dipartimento di Fisica, Universit\`a di Firenze, and INFM, 50019 Sesto Fiorentino, Italy}
\author{A. Simoni} \address{INFM and Dipartimento di Chimica,
  Universit\'a di Perugia, 06123 Perugia, Italy }

\begin{abstract}
  We determine the inter-species $s$-wave triplet scattering length
  $a_3$ for all K-Rb isotopic mixtures by measuring the elastic cross-section
  for collisions between $^{41}$K and $^{87}$Rb in different
  temperature regimes. The positive value $a_3=+163^{+57}_{-12}~a_0$
  ensures the stability of binary $^{41}$K-$^{87}$Rb Bose-Einstein
  condensates. For the fermion-boson mixture $^{40}$K-$^{87}$Rb we
  obtain a large and negative scattering length which implies an
  efficient sympathetic cooling of the fermionic species down to the
  degenerate regime.
\end{abstract}


\date{\today}
\draft \maketitle

Ultracold mixtures of alkali atoms have recently been the subject
of intensive experimental research, mainly aimed at the production
of novel degenerate quantum systems. Sympathetic cooling
\cite{ions} in a mixture represents a way to reach the degenerate
regime for species which cannot be efficiently cooled by direct
evaporation, as in the case of $^{6}$Li \cite{lithium1} and
$^{41}$K \cite{potassium}. In addition, mixing different isotopes
or atomic species allows to create ultracold Bose-Fermi systems
\cite{lithium2,ketterle}, and paves the way to the creation of
binary Bose-Einstein condensates with different atomic species and
of ultracold heteronuclear molecules.\par

In an ultracold mixture the interaction between the components is
described in terms of a single key parameter, the inter-species
$s$-wave scattering length, which for example determines the
efficiency of the sympathetic-cooling procedure. In addition,
knowledge of the scattering length is important in order to predict
macroscopic properties of these mixed systems, such as the stability
and phase separation of binary degenerate systems, or microscopic
properties, such as the occurrence of Feshbach resonances in the
inter-species collisions.  While collisions either between alkali of
the same species or between different isotopes have often been well
characterized \cite{weiner}, collisions between different alkali
species represent a still relatively unexplored field.\par

In this context, K-Rb mixtures are particularly interesting, since
they are promising candidates in the general field of Bose-Fermi
and Bose-Bose degenerate gases.  Light-assisted inelastic
collisions between various K-Rb isotopes have been investigated in
magneto-optical traps (MOTs) \cite{marcassa,jila}. Ultracold
elastic collisions have been instead studied in a magnetic trap,
but only the magnitude of the $^{41}$K-$^{87}$Rb $s$-wave triplet
scattering length has been estimated \cite{potassium}.  In this
Letter we report the results of a thorough investigation of the
collisional properties of a $^{41}$K-$^{87}$Rb mixture. By
studying the interspecies elastic collisions in a broad
temperature range, we determine the sign of the scattering length
and tighten the bounds on its magnitude. Thanks to this result, we
are able to derive by mass scaling the triplet scattering lengths
for all the other combinations of K and Rb stable isotopes and to
discuss the possibility of producing mixed K-Rb degenerate
systems.  We also determine a region of confidence for the singlet
$s$-wave scattering length of the $^{41}$K-$^{87}$Rb mixture from
inelastic decay rates of a mixed-spin sample.\par

The ultracold $^{41}$K-$^{87}$Rb mixture is produced using the
apparatus described in \cite{potassium}. In brief, about 10$^7$ K
atoms at 300~$\mu$K and 5$\times$10$^8$ Rb atoms at 100~$\mu$K are
loaded in a QUIC magnetostatic trap \cite{quic} using a
double-magneto-optical trap scheme. Both species are prepared in their
doubly-polarized spin state $|F=2,M_F=2\rangle$ thanks to an optical
pumping stage.  A gas prepared in such state has the advantage of
being immune to spin-exchange losses and is typically stable.
Evaporative cooling is then performed selectively on the Rb sample
using a microwave knife tuned to the hyperfine transition at 6.8~GHz
(the corresponding transition for K is at 254~MHz), while the K sample
is sympathetically cooled through elastic K-Rb collisions. The number
of atoms and temperature of both samples is determined simultaneously
at the end of each experimental run by absorption imaging, using two
delayed resonant light pulses after release from the magnetic trap.
Both atomic species experience the same trapping potential with
cylindrical symmetry. The axial and radial harmonic frequencies for Rb
are $\omega_{ax}=2\pi \times 16.3$~s$^{-1}$ and
$\omega_{rad}=2\pi\times 197$~s$^{-1}$ respectively, while those for K
are by a factor $(M_{Rb}/M_K)^{1/2}=1.45$ larger.\par

\begin{figure}[b]
\centerline{\includegraphics[width=8cm,clip=]{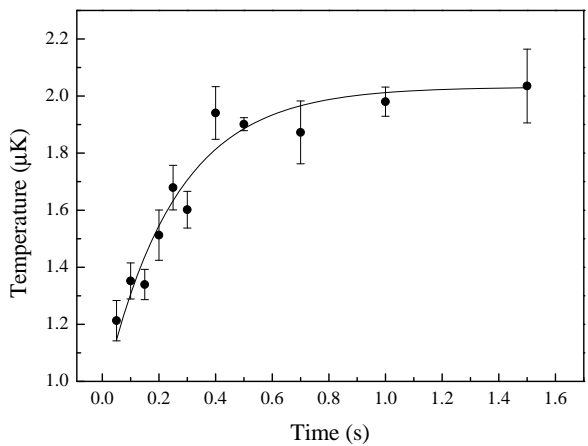}}
\caption{Evolution of the mean temperatures of the K sample
  after 100 ms of selective heating of Rb. Each datapoint is the
  average of five measurements, and the error bar displays the
  corresponding statistical fluctuation. The continuous line is the
  best fit to an exponential function; the time-constant is inversely
  proportional to the K-Rb elastic collisional rate.}
\label{therm}
\end{figure}

We have investigated the dependence of the K-Rb elastic cross-section
on temperature, by performing a series of rethermalization
measurements.  We drive the system out of thermal equilibrium and
observe the subsequent thermalization mediated by elastic collisions.
The different trap frequencies for the two species allow us to
selectively heat Rb by superimposing a 10\% modulation on the trapping
potential at twice the Rb radial frequency for typically 100~ms. As an
example we show in Fig. \ref{therm} the measured increase of
temperature of the K sample at $T\approx 1.6~\mu$K that follows
heating of Rb, together with an exponential fit to the data. The time
constant for rethermalization $\tau$ is determined from the fit. We
have checked that the aspect ratio of the K sample was constant during
the whole rethermalization which is evidence for a common temperature
of the spatial degrees of freedom.\par

\begin{figure}[b]
\centerline{\includegraphics[width=8cm,clip=]{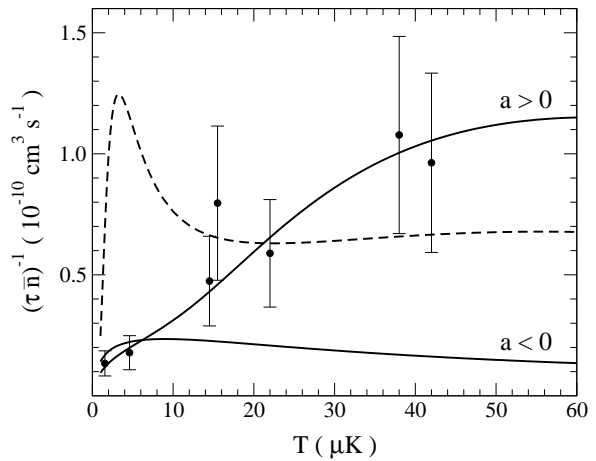}}
\caption{Dependence on temperature of the elastic collision rate
  $1/(\tau \bar{n})$. The points are the experimental data.  The solid
  lines are the best fit to the model described in the text for
  positive and negative triplet scattering lengths, corresponding to
  $a_3=163~a_0$ and $a_3=-209~a_0$, respectively.  The dashed line
  corresponds to $a_3=140~a_0$.} \label{sigma} \end{figure}

We plot in Fig.~\ref{sigma} the measured elastic-collision rate
$1/(\tau\bar{n})$ as a function of temperature in the range
$(1.6-45)~\mu$K. Here
$\bar{n}=(\frac{1}{N_K}+\frac{1}{N_{Rb}})\int n_K n_{Rb} d^3x $ is
the effective density of K and Rb atoms.
A general model to describe the thermalization between atoms of
different masses can be obtained by extending the one of
Ref.~\cite{mosk} to energy-dependent cross sections and to partial
waves higher than $s$-wave. We have checked that at our
temperatures only $s$- and $p$-wave collisions are relevant. The
thermalization time per unit atomic density mediated by $s$- and
$p$- wave collisions is
\begin{equation}
\frac{1}{\tau \bar{n}}=\frac{\xi}{2\alpha_s k_B T} \left[{\langle
\sigma_s v E \rangle} +\frac{\alpha_s}{\alpha_p} {\langle \sigma_p v E
\rangle}\right]\,,
\end{equation}
where $\alpha_s \simeq 2.7$ is the average number of $s$-wave
collisions necessary for rethermalization, $v$ is the relative
velocity of a colliding pair and $E$ is the relative collision energy.
We obtain the ratio $\frac{\alpha_s}{\alpha_p}=\frac{3}{5}$ upon
integration over all collision directions.  The factor $\xi= 4\mu/M$,
with $M$ and $\mu$ the total and the reduced mass respectively, gives
a reduction $\xi \simeq 0.87$ of the thermalization efficiency with
respect to the case of equal masses.  Finally, the averages $\langle
\cdot \rangle$ are performed on a classical Maxwell-Boltzmann
distribution of relative velocities.\par The partial cross sections
are obtained from a numerical solution of the scattering equations
using standard propagation algorithms.  The molecular Hamiltonian for
collisions of atoms with hyperfine structure includes the kinetic
energy of the relative motion, the hyperfine atomic energy and the
adiabatic Born-Oppenheimer $^1\Sigma_g^+$ and $^3\Sigma_u^+$ symmetry
potentials. The latter have been constructed by smoothly matching the
short-range {\em ab initio} potential of \cite{Lyyra} onto a
long-range analytic potential $V=V_d \pm V_{ex}/2$. Here
$V_d=-C_6/R^6-C_8/R^8-C_{10}/R^{10}$ is the dispersion potential and
$V_{ex}$ is the exchange potential. An accurate van-der-Waals
coefficient $C_6$ is taken from Ref.~\cite{Derevianko}, $C_8$ and
$C_{10}$ from \cite{Marinescu} and the analytic form of $V_{ex}$ from
\cite{Smirnov}. The molecular potential is made flexible by adding a
short-range correction to the adiabatic potentials (see
Ref.~\cite{EiteNa}). This procedure allows us to tune singlet and
triplet scattering lengths to agree with the data.  \par

The experimental data are fitted using a min-$\chi^2$ procedure. Since
collisions between atoms prepared in a doubly-polarized spin state are
single-channel collisions involving only the $^3\Sigma_u^+$ symmetry
potential, the scattering length $a_3$ is the only free parameter in
the model \cite{c6}. Our main results are shown in Fig.~\ref{sigma}.
The best agreement is obtained for a positive scattering length
$a_3=163~a_0$, while the best-fit curve for negative $a_3$ fails to
fit the experiment at high temperature (solid lines in
Fig.~\ref{sigma}). Actually, for $a_3<0$ the cross-section drops with
energy from its threshold value $\sigma_s=4\pi a_3^2$ at lower
collision energies than it does for $a_3>0$.  Moreover, the curve for
$a_3=163~a_0$ has a significant contribution from a broad $p$-wave
shape-resonance near the top of the centrifugal barrier which further
increases the rate at high $T$.  This resonance rapidly shifts at
lower energies for decreasing $a_3$ (dashed line in Fig.~\ref{sigma}).
This circumstance and the {\it absence} of observed resonant features
sets a tight lower bound $a_3 \approx 150~a_0$ on the confidence
interval for $a_3$, while the upper bound is looser.  Actually,
scattering lengths $a_3 \approx 200~a_0$ having only a minor $p$-wave
contribution still agree well with the data.\par

\begin{figure}[b]
\centerline{\includegraphics[width=8cm,clip=]{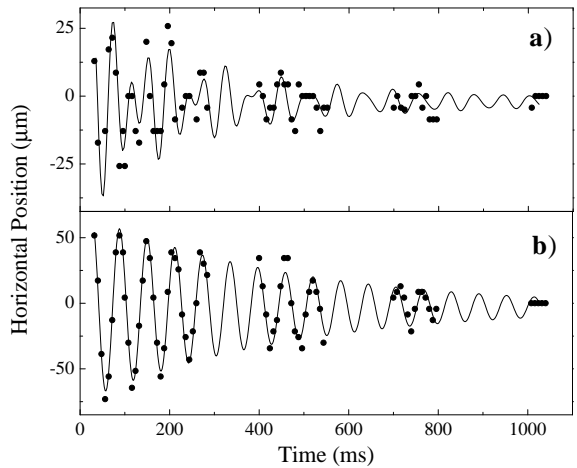}}
\caption{Dipole oscillations of the K (a) and Rb (b) clouds
  along the weak trap axis. The solid lines are a best fit to the
  model presented in the text. The faster damping and the beatings
  observed in the K motion are due to the smaller mass of the K
  sample.} \label{slosh} \end{figure}

In order to confirm our determination of the magnitude of $a_3$ we
have performed an independent measurement of the collisional
cross-section, studying the damping of dipole oscillations of a
sample composed by 8$\times 10^4$ K atoms and 1.5$\times 10^5$ Rb
atoms, at $T= 1.7 \mu$K.  Dipolar modes of the two atomic clouds
are excited by displacing the potential minimum along the weak
horizontal trap axis for about 30~ms. As shown in Fig. \ref{slosh}
the center of mass oscillations get rapidly coupled and damped by
elastic collisions. We fit the experimental data to the solutions
of the coupled equations of motion for the clouds centers of mass
\begin{eqnarray} \ddot{z}_1 &=& -\omega_1^2 z_1- \frac{4}{3}
\frac{M_2}{M} \frac{N_2}{N} \Gamma \left( \dot{z}_1 - \dot{z}_2 \right)
\nonumber \\ \ddot{z}_2 &=& -\omega_2^2 z_2 + \frac{4}{3} \frac{M_1}{M}
\frac{N_1}{N} \Gamma \left( \dot{z}_1 - \dot{z}_2 \right)\,,
\label{sloshing} \end{eqnarray}
where the labels 1 and 2 denote K and Rb respectively, $M=M_1+M_2$,
$N=N_1+N_2$, $\Gamma = {\bar{n}} \sigma_s v$ is the $s$-wave collision
rate, $v$ is the rms relative velocity, and $\bar{n}$ is the atoms
density defined above.  The low temperature at which the experiment is
performed ensures that $s$-wave collisions dominate.  Our system
appears to be in the collisionless regime (
$\Gamma<<\omega_1,\omega_2$), where the position of each center of
mass is the sum of two damped sinusoids at the bare frequencies of K
and Rb.  Our best fit, shown in Fig.~\ref{slosh}, well reproduces the
experimental observation.  From the fitted collision rate $\Gamma=
5.2(7)$~s$^{-1}$ we determine the magnitude of triplet scattering
length $|a_3|=170^{+35}_{-35}~a_0$, where the uncertainty is dominated
by that on the atom number. This value for $|a_3|$ is in good
agreement with the one previously derived from thermalization
measurements.\par

\begin{figure}[b]
\centerline{\includegraphics[width=7.5cm,clip=]{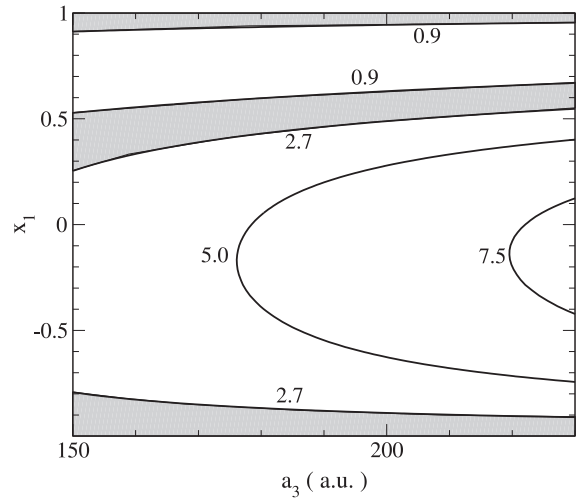}}
\caption{Contour plot of the calculated inelastic collisional rate
$G$ in units of $10^{- 11}$~cm$^3$~s$^{-1}$ {\it vs} the triplet
scattering length $a_3$ and the
  reduced singlet scattering length $x_1=2/\pi \arctan(a_1/a_{sc})$,
  where $a_{sc}=72~a_0$. The shaded regions correspond to the rate
  determined in the experiment.}
\label{inelastic} \end{figure}

We determine the singlet K-Rb scattering length $a_1$, by measuring
the trap loss for collision between K atoms in the hyperfine state
$|2,2\rangle$ and Rb atoms in $|2,1\rangle$.  Our sample is initially
composed by a small number of K atoms (3$\times 10^4$) and a larger
number of Rb atoms (1.2$\times 10^5$), both in $|2,2\rangle$, at a
temperature $T=1.8~\mu$K. We then transfer a small fraction (20\%) of
both species in $|2,1\rangle$ using a radio-frequency sweep and we
study the subsequent decay of K, which takes place on a timescale much
shorter than the lifetime of the doubly spin-polarized mixture. In
these experimental conditions K atoms in $|2,1\rangle$ and Rb atoms in
$|2,2\rangle$ do not contribute significantly to loss of K atoms from
the trap. The time evolution of K-atoms number is given to a good
approximation by the solution of the single rate equation
$\dot{n}^{K}(t)=-G\, n^{K}_{|2,2\rangle}(t)\, n^{Rb}_{|2,1\rangle}(t)$
with the constraint $N^{K}_{|2,2\rangle}(t)-N^{Rb}_{|2,1\rangle}(t)=$
const. Here $n$ is the spatial density, $N$ the atom number and $G$ is
the rate constant for inelastic collisions between K in $|2,2\rangle$
and Rb in $|2,1\rangle$ \cite{rate}.  Our collision model agrees well
with the experimental data for an inelastic collision rate
$G$=1.8(9)$\times 10^{- 11}$~cm$^3$~s$^{-1}$. A comparison between the
measured inelastic rate and the numerical calculation (shown in
Fig.~\ref{inelastic}) allows us to determine the region of confidence
for the singlet scattering length $a_1$, given our interval of $a_3$
values. This comparison singles out a region of positive singlet
scattering length, $a_1>30~a_0$, excluding a broad region of strong
suppression of inelastic processes centered at $a_1 \sim a_3$.  In
alternative, the large negative singlet scattering length $a_1 <
-210~a_0$ is also compatible with the data.  This result represents a
first characterization of the singlet K-Rb interaction, and can
provide a useful guidance for further collisional studies, in
particular for the investigation of Feshbach resonances in selected
spin-states.\par

We have also determined the triplet $s$-wave scattering lengths
for all the other pairs of K and Rb isotopes.  It must be remarked
that our analysis for $^{41}$K-$^{87}$Rb is weakly dependent on
the number $N_b$ of bound states supported by the ABO potentials.
In contrast, mass scaling from one isotope to the others depends
more strongly on $N_b$. Tab. \ref{table} summarizes our results
for the number of bound states $N_b=32$ supported by the nominal
triplet potential. Calculations that allow for a realistic $(\pm
2)$ uncertainty in $N_b$ do not qualitatively change the results
of Tab.~\ref{table}. With this proviso, we discuss below the
consequences of these results.
\begin{table}[t]
\begin{center}
\caption{Triplet $s$-wave scattering lengths (in a.u.) for
collisions between K and Rb isotopes computed using the nominal
triplet potential.} \label{table} \vskip 12pt
\begin{tabular}{c c c c}  & $^{39}$K  & $^{40}$K & $^{41}$K \\ \hline \\
 $^{85}$Rb &58$^{+14}_{-6}$&-38$^{+37}_{-17}$&329$^{+1000}_{-55}$\\
$^{87}$Rb &31$^{+16}_{-6}$&-261$^{+170}_{-159}$&163$^{+57}_{-12}$
\end{tabular}
\end{center}
\end{table}
The repulsive character of the $^{41}$K-$^{87}$Rb triplet
interaction indicates that it is possible to form a stable binary
Bose-Einstein condensate.  Indeed, in case of a {\it negative}
inter-species scattering length much larger in magnitude than the
single-species ones \cite{notaa3}, the mean-field interaction
would lead to a collapse of the condensates
\cite{binarybecs,notabec}.  Of great interest is also the
$^{40}$K-$^{87}$Rb pair, for which we find a negative and probably
large inter-species scattering length. Therefore, at least in the
absence of unfortunately placed zeros in the elastic cross
section, sympathetic cooling of fermionic K with $^{87}$Rb can be
expected to be highly efficient.  Moreover, the attractive
character of the interaction should prevent phase separation of
the two components once the degenerate regime has been reached
\cite{moelmer}, thus insuring an efficient thermalization even in
this regime. This represent a qualitatively different situation
with respect to the other known Bose-Fermi mixture, $^6$Li-$^7$Li,
for which the interspecies scattering length is positive
\cite{lithium3}. We note that among the pairs containing
$^{85}$Rb, which however does not seem to be an easy partner for
sympathetic cooling \cite{rb85}, only the $^{41}$K-$^{85}$Rb pair
shows a significantly large interaction. Finally, the mixture
$^{39}$K-$^{87}$Rb is not particulary appealing, since the
interspecies scattering length is expected to be relatively small
and, moreover, the formation of a stable Bose-Einstein condensate
of $^{39}$K with a large number of atoms is prevented by the
negative sign of its own scattering length \cite{wang}.\par

In conclusion we have drawn a picture of the triplet interaction
between all the K-Rb isotopes pairs. Our study paves the way to
the production of novel mixtures of degenerate atomic gases, such
as Bose-Bose systems with large repulsive interaction and
Bose-Fermi systems with attractive interactions. Moreover, it
provides important information for predicting the occurrence of
magnetically-tunable Feshbach resonances. These resonances would
represent a tool to control the inter-species interactions and to
drive the production of ultracold heteronuclear molecules.
\par
We are grateful to P. S. Julienne, F. H. Mies, E.  Tiesinga, and C. J.
Williams for the ground-state collisions code and several useful
discussions. This work was supported by MURST, by ECC under the
Contract HPRICT1999-00111, and by INFM, PRA "Photonmatter".


\begin{thebibliography}{99}
\bibitem{c6} We have checked that a $\pm$1\% uncertainty in the highly
accurate $C_6$ coefficient adopted in this work does not affect the results of the paper.
\bibitem{rate} The only relevant inelastic processes involved are $|2,1\rangle+|2,2\rangle \to
 |2,2\rangle+|1,1\rangle$ and $|2,1\rangle+|2,2\rangle \to
 |1,1\rangle+|2,2\rangle$.
\bibitem{notaa3} The triplet scattering lengths for $^{41}$K and $^{87}$Rb are $a_3=60~a_0$ \cite{wang} and $a_3=99~a_0$ \cite{verhaar}, respectively.
\end{thebibliography}
\end{document}